\documentclass[
  aps,
  prl,
  twoside,
  twocolumn,
  showpacs,
  floatfix,
  10pt,
  superscriptaddress,
]{revtex4-1}

\usepackage{graphicx}
\usepackage{amsmath}

\newcommand{\fig}[1]{Fig.~\ref{#1}}

\usepackage{color}

\begin{document}

\title{
  Low-temperature criticality of martensitic transformations of Cu nanoprecipitates in $\alpha$-Fe
}

\author{Paul Erhart}
\email{erhart@chalmers.se}
\affiliation{
  Chalmers University of Technology,
  Department of Applied Physics,
  Gothenburg, Sweden
}
\affiliation{
  Lawrence Livermore National Laboratory,
  Condensed Matter and Materials Division,
  Livermore, California, 94551, USA
}
\author{Babak Sadigh}
\email{sadigh1@llnl.gov}
\affiliation{
  Lawrence Livermore National Laboratory,
  Condensed Matter and Materials Division,
  Livermore, California, 94551, USA
}

\begin{abstract}
 Nanoprecipitates form during nucleation of multiphase equilibria in phase segregating multicomponent systems. In spite of their ubiquity, their size-dependent physical chemistry, in particular at the boundary between phases with incompatible topologies, is still rather arcane. Here we use extensive atomistic simulations to map out the size--temperature phase diagram of Cu nanoprecipitates in $\alpha$-Fe. The growing precipitates undergo martensitic transformations from the body-centered cubic (BCC) phase to multiply-twinned 9R structures. At high temperatures, the transitions exhibit strong first-order character and prominent hysteresis. Upon cooling the discontinuities become less pronounced and the transitions occur at ever smaller cluster sizes. Below 300\,K the hysteresis vanishes while the transition remains discontinuous with a finite but diminishing latent heat. This unusual size-temperature phase diagram results from the entropy generated by the soft modes of the BCC-Cu phase, which are stabilized through confinement by the $\alpha$-Fe lattice.
\end{abstract}


\maketitle

\begin{figure*}
  \centering
  \includegraphics[width=0.95\linewidth]{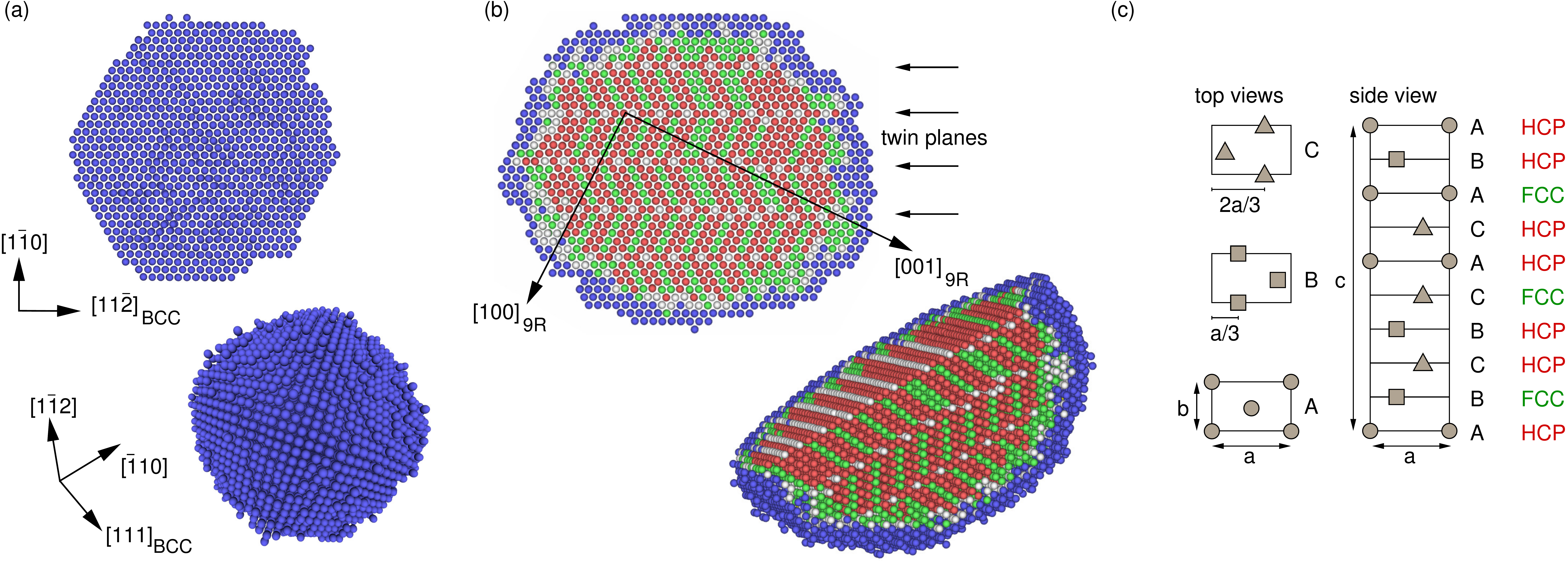}
  \caption{
    (a,b) Cross sectional (top) and perspective (bottom) views of (a) BCC and (b) twinned 9R Cu precipitates at 700\,K. The precipitates contain approximately 15,000 and 36,000 atoms, respectively. The 9R precipitate in (b) exhibits a pronounced BCC wetting layer.
    (c) Schematic of 9R structure \cite{OthJenSmi91, OthJenSmi94} which appears as a periodic sequence of two HCP atoms and one FCC atom along the $[001]_{\text{9R}}$ direction when using the Ackland-Jones parameter.
  }
  \label{fig:viz_prec}
\end{figure*}

The Fe--Cu system is a prototypical structurally mismatched system with very small solubility of Cu in Fe. At low temperatures Fe adopts the body-centered cubic (BCC) phase, while Cu prefers the face-centered cubic (FCC) phase with BCC-Cu being mechanically unstable. Phase segregation in Fe--Cu alloys also has been and continues to be the subject of considerable technological interest since Cu impurities contribute to high-temperature embrittlement of ferritic steels \cite{OdeLuc01}. Based on a series of careful experimental studies three stages in the transformation of Cu precipitates have been identified \cite{OthJenSmi91, OthJenSmi94, LeeKimKim07, PhyForEng92, LudFarPed98, BlaAck01, Bou01}: ({\it i}) Cu precipitates nucleate in the BCC structure of the host lattice, ({\it ii}) as they grow they undergo a martensitic phase transition to a multiply-twinned 9R structure, which ({\it iii}) eventually transforms into FCC. In this Letter using a novel combination of molecular dynamics (MD) and Monte Carlo (MC) simulations as well as empirical interatomic potentials, we develop a comprehensive picture of the correlation between precipitate structure and size as a function of temperature for the first two stages of the transformation process. 

The Fe--Cu system was modeled using semi-empirical interatomic potentials \cite{MisMehPap01, MenHanSro03, PasMal07} that provide an accurate description of the alloy phase diagram and yield very good agreement with density-functional theory calculations for the energies of small Cu clusters in BCC-Fe. Simulations were carried out using the hybrid MC/MD algorithm introduced in \cite{SadErhStu12, SadErh12} implemented in the massively parallel MD code \textsc{lammps} \cite{Pli95}.
Orthorhombic simulation cells were employed with cell vectors oriented along $[111]$, $[1\bar{1}0]$, and $[11\bar{2}]$. Most simulations used simulation cells containing $42\times 52\times 45$ unit cells (786,240 atoms) with additional simulations of smaller precipitates at low temperatures using $31\times 38\times 33$ unit cells (310,992 atoms). Periodic boundary conditions were applied in all directions, and pressure and temperature were controlled using Nos\'e-Hoover thermostat and barostat \cite{FreSmi01}. Simulations were run for at least $1.6\times 10^6$ and up to $4\times 10^6$ MD steps using a timestep of 2.5\,fs. Every 20 steps the MD simulation was interrupted to carry out a number of VC-SGC MC trial moves corresponding to 10\%\ of a full MC sweep. Each simulation run thus comprises between 8,000 and 20,000 attempts per particle to swap the atom type. The atomic structures were analyzed using the Ackland-Jones parameter \cite{AckJon06}, which uses bond angle distributions to classify local environments as BCC, FCC, or HCP. The accuracy of this analysis was enhanced by position-averaging over 800 MD steps. Precipitate structures have been rendered using \textsc{OVITO} \cite{Stu10}.

Figure~\ref{fig:viz_prec} exhibits precipitate structures observed in simulations at 700\,K. Initially, Cu precipitates are isostructural with the surrounding BCC-Fe matrix, as shown exemplarily in Figure~\ref{fig:viz_prec}(a) by a precipitate containing approximately 15,000 atoms. Even larger precipitates undergo a structural transformation from the BCC phase to a multiply-twinned 9R structure as illustrated in \fig{fig:viz_prec}(b) where a cluster containing about 36,000 Cu atoms is shown. The 9R phase is a closed-packed (CP) lattice that differs from FCC through its stacking sequence of the CP planes, as shown on the right-hand side of \fig{fig:viz_prec}(c). Using the Ackland-Jones parameter for structure identification (see Method section), we obtain sequences of FCC and hexagonal close-packed (HCP) atoms in \fig{fig:viz_prec}(b) that are in accord with the ideal 9R structure, c.f. \fig{fig:viz_prec}(c). In addition, we observe stacking faults as well as several twins as indicated by arrows in \fig{fig:viz_prec}(b). These results closely resemble the TEM micrographs of 9R Cu precipitates in $\alpha$-Fe \cite{OthJenSmi91, OthJenSmi94, LeeKimKim07}. Further analysis reveals that also parameters such as twin spacing and misorientation angles agree very well with experiments.

\begin{figure}[th]
  \centering
  \includegraphics[scale=0.62]{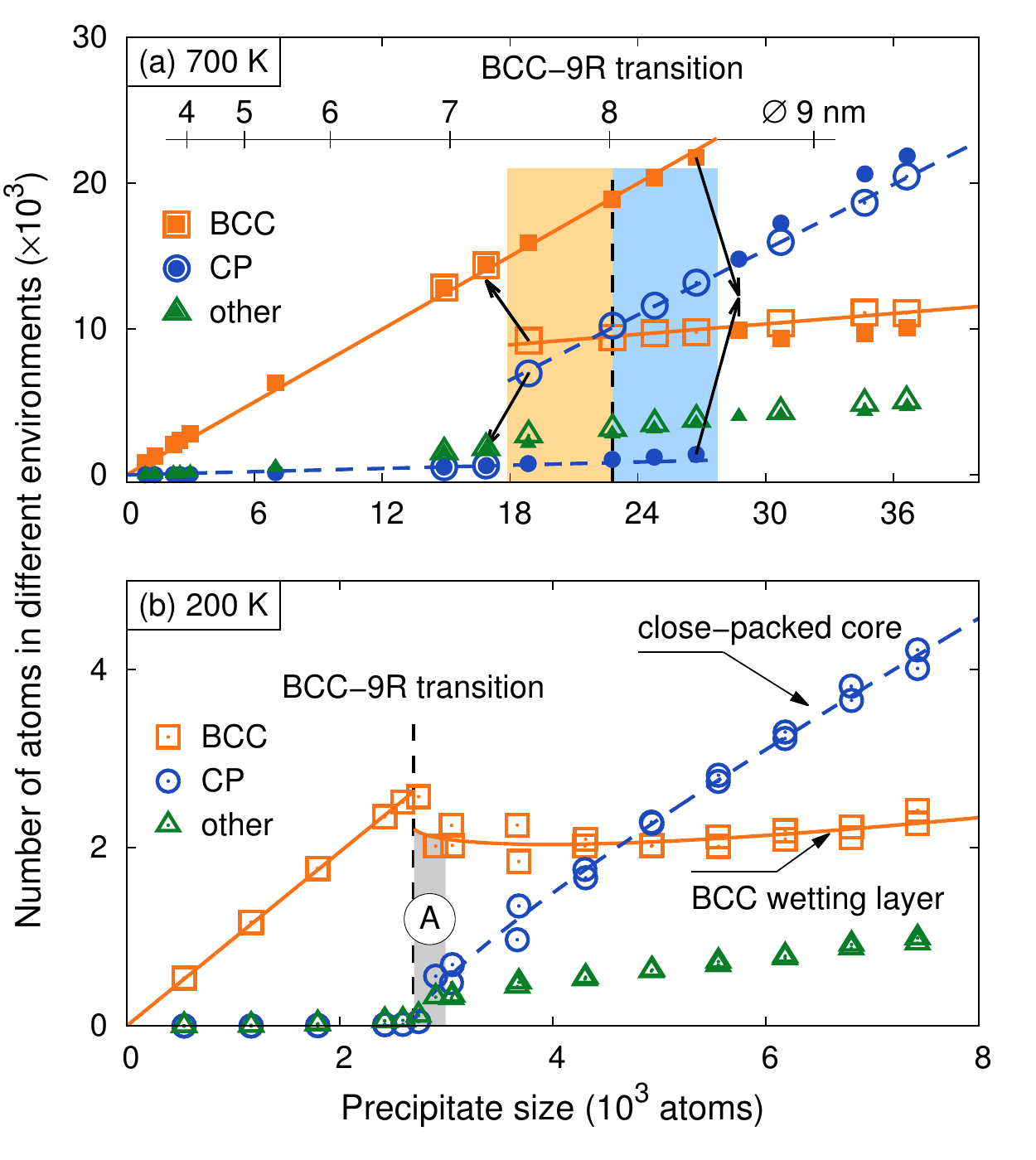}
  \caption{
    Number of atoms in different local environments at (a) 700\,K and (b) 200\,K. Filled and open symbols in (a) show results from forward (BCC$\rightarrow$9R) and backward (9R$\rightarrow$BCC) simulations, respectively. The BCC--9R transition exhibits a hysteresis at higher temperatures that vanishes as temperature is lowered.
  }
  \label{fig:trans_struct}
\end{figure}

In order to map out the size--temperature phase diagram of Cu nanoprecipitates, we record the number of Cu atoms in CP/BCC local environments as a function of precipitate size at various temperatures. The open symbols in \fig{fig:trans_struct}(a) show the evolution of these populations with increasing precipitate size at 700\,K. Precipitates exceeding 28,000 atoms are found to undergo a {\em discontinuous} structural transition as they transform from BCC to multiply-twinned 9R structures. To study the reverse transition, we start from a 9R precipitate obtained from an earlier simulation and gradually reduce the Cu content in the simulation. The thus obtained data are shown by the filled symbols in \fig{fig:trans_struct}(a) demonstrating that at approximately 18,000 atoms precipitates revert to the BCC structure. There is thus a pronounced hysteresis between forward (BCC$\rightarrow$9R) and backward (9R$\rightarrow$BCC) transitions at 700\,K extending from about 18,000 atoms to 28,000 atoms.

\begin{figure}
  \centering
  \includegraphics[scale=0.62]{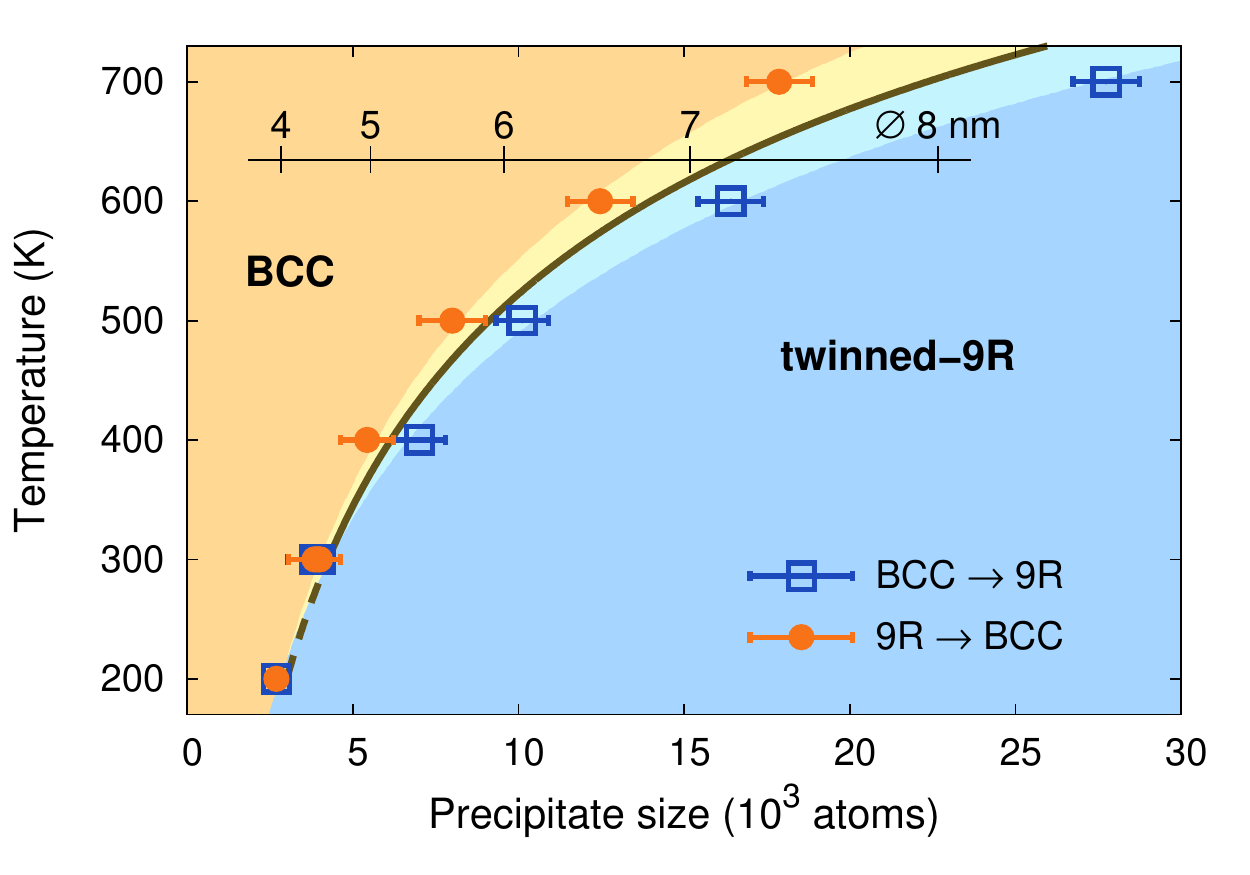}
  \caption{
    Diagram representing the structure of Cu precipitates as a function of temperature and size. The light colored regions next to the phase transition line delineate the extent of the hysteresis. The hysteretic critical point is located along the dashed line between 200 and 300\,K.
  }
  \label{fig:temp_trans}
\end{figure}

The above analysis was repeated for six temperatures between 200 and 700\,K, the results of which lead to the size--temperature phase diagram depicted in \fig{fig:temp_trans}. It demonstrates that the critical cluster size for the BCC--9R transition is strongly temperature dependent increasing from approximately 2,700 atoms (3.9\,nm diameter) at 200\,K to about 22,000 atoms (7.9\,nm) at 700\,K. The transition is decidedly first-order at high temperatures with a pronounced hysteresis that vanishes below 300\,K, whereupon the transition becomes {\em nearly continuous}. This is depicted in \fig{fig:trans_struct}(b), where the numbers of BCC and CP atoms change around the critical size of 2,700 atoms reversibly with negligible discontinuity. 
A more detailed inspection of the precipitate structure at 200\,K (see movies and snapshots in Supplementary Material) shows that precipitates below the transition point (2,700 atoms) have completely adopted a BCC lattice structure while precipitates containing more than 3,000 atoms exhibit multiply-twinned 9R structures as described above. In the intermediate size range [region A in \fig{fig:trans_struct}(b)] the atoms that constitute the core of the precipitate rapidly fluctuate between different CP motifs. These fluctuations are absent at higher temperatures where the BCC--9R transition is sharp.

It is important to note that irrespective of temperature even after the transition to 9R the precipitates still contain a substantial number of BCC atoms, which grows with precipitate size, albeit much slower than the number of CP atoms. This observation reflects the existence of a BCC-Cu wetting layer between the BCC-Fe matrix and the CP core of the Cu precipitate. It can be as thick as 4 atomic layers at the transition point at 200\,K, while it only amounts to about 1 to 2 atomic layers for very large precipitates. The BCC-Cu wetting layer is also clearly visible in \fig{fig:viz_prec}(b). In the following, we will see that the wetting layer plays a crucial role in the BCC--9R transition, particularly at low temperatures. 

\begin{figure}
  \centering
  \includegraphics[scale=0.62]{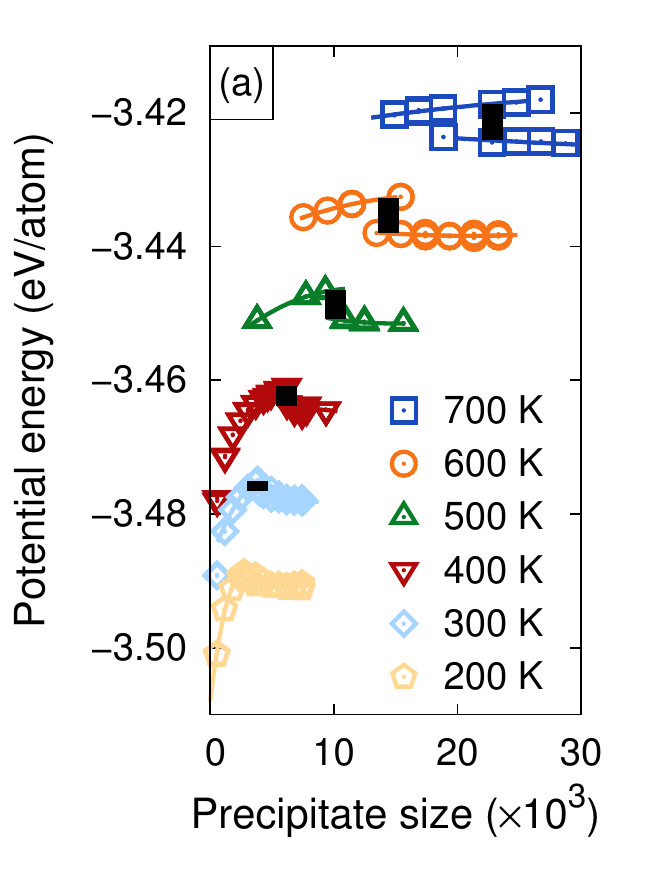}
  \includegraphics[scale=0.62]{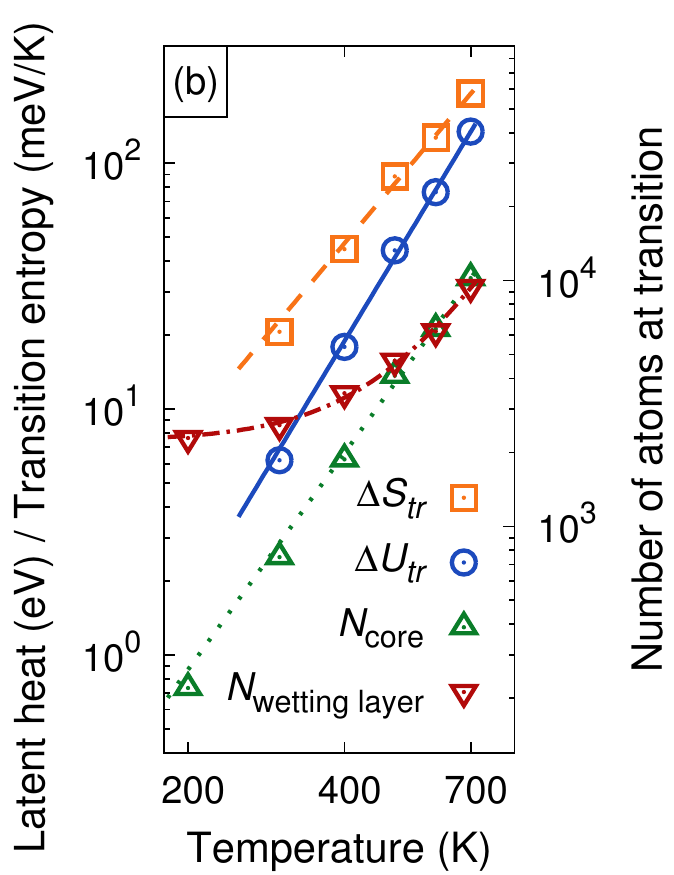}
  \caption{
    (a) Potential energy of atoms in Cu precipitates as a function of precipitate size for different temperatures. The vertical black bars represent the latent heat $\Delta U_{tr}$ associated with the transformation, which is plotted in (b) as a function of temperature alongside the entropy change at the transition ($\Delta S_{tr} = \Delta U_{tr} / T_{tr}$). The upward (downward) pointing triangles show the number of core (wetting layer) atoms at the transition. Lines are guides to the eye.
  }
  \label{fig:epot_trans}
\end{figure}

An important signature of first-order transitions is the release of latent heat, which is absent in continuous transitions. To conclusively resolve the character of the observed transitions we have extracted the internal energy density of the Cu atoms in the precipitates as a function of precipitate size at various temperatures as shown in \fig{fig:epot_trans}(a). At high temperatures the data exhibit pronounced discontinuities when the higher-energy BCC phase transforms to the more stable 9R phase. As the temperature is lowered the discontinuity in the potential energy is reduced until it apparently vanishes between 200 and 300\,K, see \fig{fig:epot_trans}(b). If true, this implies that a critical point exists in this temperature range, below which the BCC--9R transition occurs continuously. However, the observation of a continuous transition contradicts bulk thermodynamics, according to which the distinct lattice symmetries require the BCC--9R transformation to be first order. Can this rule be violated for precipitate sizes as astoundingly large as 2,500 to 3,000 atoms?

\begin{figure*}
  \centering
  \includegraphics[scale=0.6]{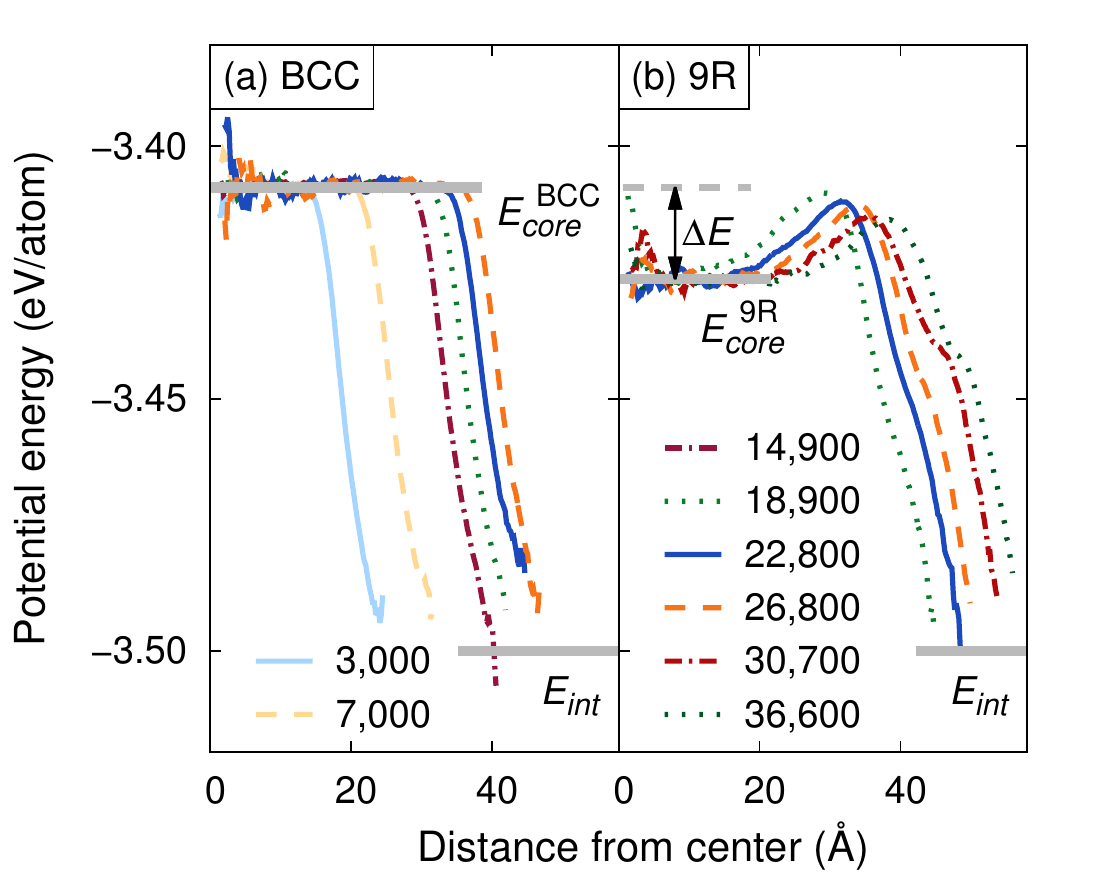}
  \includegraphics[scale=0.6]{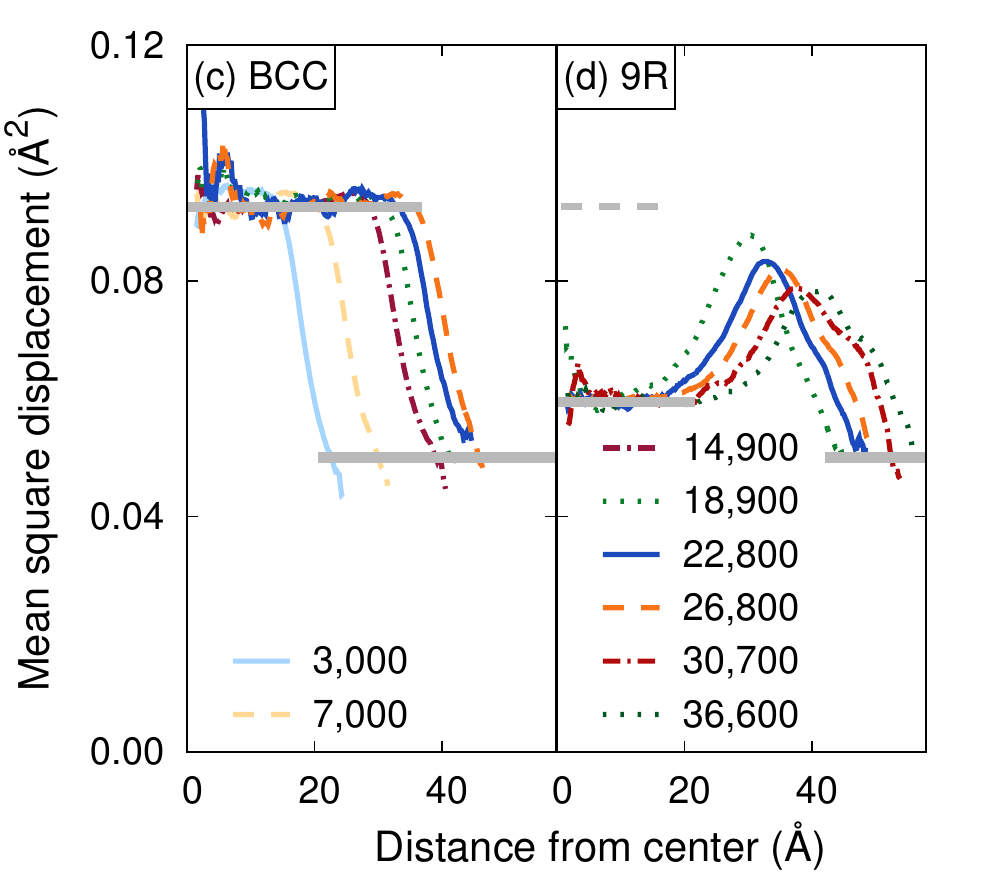}
  \includegraphics[scale=0.6]{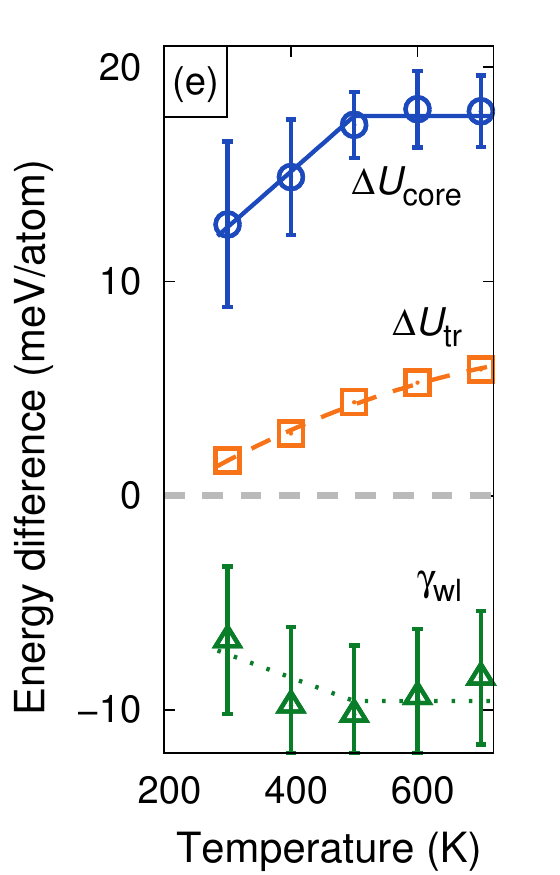}
  \caption{
    (a,b) Potential energy and (c,d) mean square displacement as a function of distance from precipitate center for (a,c) BCC and (b,d) 9R precipitates at 700\,K.
    (e) Latent heat $\Delta U_{tr}$ [from \fig{fig:epot_trans}(b)] as well as core $\Delta U_\text{core}$ and wetting layer $\gamma_\text{wl}$ energy differences between BCC and 9R precipitates obtained by analyzing PE profiles.
  }
  \label{fig:profiles}
\end{figure*}

The key to answering this question lies in the heterogeneous distribution of energy and entropy inside Cu precipitates. For the purpose of this discussion we estimate entropy by employing the Debye model for the vibrational spectrum, which provides a simple relation between entropy and mean square displacement (MSD).
\footnote{By parametrizing effective Debye models for the vibrational spectra of the two phases of interest, one can approximate their entropy difference as $\Delta S = \frac{1}{2} k_B \ln\left( \left<\Delta r_2^2\right> / \left<\Delta r_1^2\right> \right)$, where $\left<\Delta r_i^2\right>$ is the MSD of phase $i$.}
Figure~\ref{fig:profiles}(a,b) shows the average atomic potential energies (PE), and \fig{fig:profiles}(c,d) depicts the MSD as a function of distance to the centers of precipitates at 700\,K for several BCC and 9R precipitates. The PE and MSD profiles of the BCC precipitates have similar shapes with a constant region in the core that remains unchanged as the precipitates grow. The wetting layer is identified as the region outside the core where both the PE and the MSD of the Cu atoms drop steeply. This confirms that the vibrational entropy of the core region of BCC precipitates is much larger than the wetting layer entropy. It leads to the stabilization of BCC precipitates at elevated temperatures and derives from the vibrational modes of bulk BCC-Cu with imaginary frequencies that are stabilized through confinement by the Fe matrix.

Inspection of the PE and MSD profiles of the 9R precipitates in \fig{fig:profiles}(b,d) reveals a more complex behavior than in the case of the BCC precipitates. While the wetting layer seems to be quite similar in size, energy and entropy to the BCC precipitates, the core now consists of two regions: ({\it i}) the inner core with uniform energy and entropy, and ({\it ii}) the outer core, which constitutes the interface region between the twinned 9R inner core and the wetting layer. It is strained to accommodate the misorientation between the inner-core 9R lattice and the BCC-Fe matrix. In this region, both energy and MSD increase but never exceed the values of the core of the BCC precipitates. Hence the 9R core always contributes positively to the latent heat. However, the total latent heat $\Delta U_{tr}$ generated by the structural transformation is always smaller than the core contribution since the wetting layer energy is always slightly lower in BCC than in 9R precipitates. The total latent heat and its associated entropy $\Delta S_{tr}=\Delta U_{tr}/T$ are shown in Fig.~\ref{fig:epot_trans}(b). Both decrease dramatically upon cooling mainly because at lower temperatures transitions involve much fewer atoms as they occur at much smaller cluster sizes. Furthermore at lower temperatures the latent heat per atom drops significantly [see \fig{fig:profiles}(e)] due to the fact that in smaller precipitates the core region, which provides the positive contribution to the latent heat, shrinks while the compensating wetting layer thickens. This is supported by the number of atoms in the core and the wetting layer regions at the transition as depicted in Fig.~\ref{fig:epot_trans}(b). 

Nevertheless based on the reported temperature dependence of the entropy of transformation in Fig.~\ref{fig:epot_trans}(b), one expects the BCC--9R transformation to involve a diminishing but finite latent heat at temperatures clearly below the critical temperature ($\approx 300\,\text{K}$) at which the hysteresis vanishes. Note that at this temperature, as many as 500 Cu atoms form the core of the 9R precipitate and thus directly participate in the transformation, see Fig.~\ref{fig:epot_trans}(b). These observations thus support the existence of a dynamic critical point in the BCC--9R transition of Cu nanoprecipitates. At this temperature, the nucleation barrier between the two phases disappears. The microscopic origin of this behavior lies in the mechanical instability of bulk BCC-Cu on the one hand and the stabilization of a BCC wetting layer with stiff phonon modes at the interface with the BCC-Fe host lattice on the other.

It is important to note that while our observations cannot prove the existence of a thermodynamic critical point (TCP) in the BCC--9R transition of Cu nanoprecipitates, they suggest that the phase diagrams of nanoprecipitates may contain TCPs that have no counterpart in the corresponding bulk thermodynamic limit. It is of course well-known that the phase diagrams of nanoparticular systems deviate from the bulk due to confinement. In particular the loci of phase boundaries become size dependent and thus phase transition temperatures and pressures are modified for liquid-gas \cite{BalEva89}, liquid--solid \cite{Tak54, Sam71, GolEchAli92, CouJes77, BufBor76} as well as solid--solid transformations \cite{TolAli94, CheHerJoh97, JacZazSch01, TolAli95a, SchEic94, FrePay96, LopHayBoa02, WanWenHof10, XioQiHua11, AbdCraLam12}. Furthermore, dynamic critical points leading to vanishing hysteresis have been observed to precede the TCP in liquid-gas transitions \cite{BalEva89, NeiVis06}. In contrast, one might expect the first-order character of transitions between phases with distinct symmetries to be conserved. The presence of a TCP in BCC--9R transformations of Cu nanoprecipitates violates this expectation. It is caused by the heterogeneous nature of nanoprecipitates, where the total latent heat can vanish with the core and wetting layer regions exhibiting compensating discontinuities. This heterogeneity is best illustrated in \fig{fig:profiles}(e), which shows the total latent heat per atom, as well as the inner core and wetting layer energy differences between BCC and 9R precipitates. For temperatures above 500\,K, the contribution to the latent heat from the inner core region reaches its asymptotic limit. Below this temperature, this contribution is slowly reduced as the inner core region is gradually shrinking and becoming strained. Figure~\ref{fig:profiles}(e) demonstrates that the latent heat per atom can nearly vanish while significant differences in the energetics of BCC and 9R precipitates remain. This points to the fact that the balance between wetting layer and core energies is fundamental to the low-temperature criticality in this system. Furthermore, if a TCP existed, higher order derivatives of the free energy would become discontinuous rather than diverge.

The phenomenology described in this work not only has direct consequences for our understanding of the Fe--Cu system but transpires to many more materials. Based on the analysis presented here one can anticipate the occurrence of similar effects in other immiscible systems, whenever the minority phase is unstable in the lattice structure of the host.

\section*{Acknowledgments}
Parts of this work were performed under the auspices of the U.S. Department of Energy by LLNL under Contract DE-AC52-07NA27344. P.E. acknowledges funding from the Swedish Research Council in the form of a Young Researcher grant and the {\em Area of Advance -- Materials Science} at Chalmers. B.S. acknowledges funding from the DOE-NE NEAMS program. Com\-puter time allocations by the Swedish National Infrastructure for Computing are gratefully acknowledged.


%

\end{document}